\documentclass[review]{elsarticle}

\usepackage{hyperref}

\journal{Journal of Computer and System Sciences}
\usepackage{amsmath}
\usepackage{amssymb}
\usepackage{color}
\usepackage{graphicx}
\usepackage{algorithmic}

\usepackage{algorithm}

\def\>{\ensuremath{\rangle}}
\def\<{\ensuremath{\langle}}

\def\-{\ensuremath{\textrm{-}}}

\def\h{\ensuremath{\mathcal{H}}}
\def\p{\ensuremath{\mathcal{P}}}

\def\g{\ensuremath{\mathcal{G}}}

\def\r{\ensuremath{\mathcal{R}}}
\def\R{\ensuremath{\mathfrak{R}}}
\def\f{\ensuremath{\mathcal{F}}}

\def\u{\ensuremath{\mathcal{U}}}
\def\k{\ensuremath{\mathcal{K}}}

\def\t{\ensuremath{\mathcal{T}}}
\def\u{\ensuremath{\mathcal{U}}}

\def\x{\ensuremath{\mathcal{X}}}
\def\y{\ensuremath{\mathcal{Y}}}

\def\v{\ensuremath{\mathcal{V}}}

\def\b{\ensuremath{\mathcal{B}}}

\def\e{\ensuremath{\mathcal{E}}}
\def\f{\ensuremath{\mathcal{F}}}

\newcommand{\supp}[1]{\ensuremath{\lceil{#1}\rceil}}

\def\<{\langle}
\def\>{\rangle}

\def\k{\mathcal{K}}
\def\E{\mathcal{E}}
\def\G{\mathcal{G}}
\def\H{\mathcal{H}}
\def\R{\mathcal{R}}
\def\supp{\textrm{supp}}
\def\SC{\textrm{SC}}

\def\supp{\mathit{supp}}
\newtheorem{theorem}{Theorem}

\newtheorem{corollary}{Corollary}
\newtheorem{lemma}{Lemma}

\newtheorem{definition}{Definition}
\newtheorem{example}{Example}

\begin{document}

\begin{frontmatter}

\title{Decomposition of Quantum Markov Chains and Its
Applications}

\author[mymainaddress]{Ji Guan\corref{mycorrespondingauthor}}
\cortext[mycorrespondingauthor]{Corresponding author.}
\ead{Ji.Guan@student.uts.edu.au}

\author[mymainaddress]{Yuan Feng}
\ead{Yuan.Feng@uts.edu.au}

\author[mymainaddress,mysecondaryaddress,mythirdaddress]{Mingsheng Ying}
\ead{Mingsheng.Ying@uts.edu.au}

\address[mymainaddress]{Center for Quantum Software and Information,
University of Technology Sydney, NSW 2007, Australia}
\address[mysecondaryaddress]{Institute of Software, Chinese Academy of Sciences, Beijing 100190, China}
\address[mythirdaddress]{Department of Computer Science and Technology, Tsinghua University, Beijing 100084, China}
 
\begin{abstract}
Markov chains have been widely employed as a fundamental model in the studies of probabilistic and stochastic communicating and concurrent systems. It is well-understood that decomposition techniques play a key role in reachability analysis and model-checking of Markov chains. (Discrete-time) quantum Markov chains have been introduced as a model of quantum communicating systems \cite{wolf2012quantum} and also a semantic model of quantum programs \cite{ying2013verification}. The BSCC  (Bottom Strongly Connected Component) and stationary coherence decompositions of quantum Markov chains were introduced in {\color{black}\cite{umanita2006classification,ying2013reachability,baumgartner2012structures}}. This paper presents a new decomposition technique, namely periodic decomposition, for quantum Markov chains. We further establish a limit theorem for them. As an application, an algorithm to find a maximum dimensional noiseless subsystem of a quantum communicating system is given using decomposition techniques of quantum Markov chains.
 \end{abstract}
 \end{frontmatter}

\section{Introduction}
A Markov chain is a random process with the property that, conditional on its present value, the future is independent of the past. Markov chains have been used as statistical models of real-world processes in a wide range of fields \cite{grimmett2001probability}. They have been extensively employed as a fundamental model of probabilistic and stochastic communicating and concurrent systems. In particular, various algorithmic analysis and model checking techniques have been developed for them in the last three decades (see for example \cite{baier:model-checking}, Chapter 10). 

The notion of Markov chain, when properly generalized to the quantum world, provides a potential paradigm for modeling the evolution of quantum systems. Continuous-time quantum Markov processes have been intensively studied in mathematical physics for many years, and achieved several discoveries of fundamental importance~{\color{black}\cite{davies1969quantum,davies1970quantum,lindblad1976generators,fagnola2001existence}}. Discrete-time quantum Markov chains {\color{black}with  finite dimensions} were first introduced as a model of quantum communicating systems \cite{wolf2012quantum}. 
A special class of quantum Markov chains, namely quantum walks, has been successfully used in the design of quantum algorithms (see \cite{ambainis2003quantum,kempe2003quantum} for a survey of this line of research). More recently, discrete-time quantum Markov chains were introduced by the authors and their collaborators \cite{ying2013verification,yu2012reachability,ying2013reachability} as a semantic model for the purpose of verification and termination analysis of quantum programs.

It is well-known that decomposition techniques, e.g. the BSCC (Bottom Strongly Connected Component) decomposition, are a key to reachability analysis and model checking of Markov chains (see for example, \cite{baier:model-checking}, Section 10.1.2). In \cite{ying2013reachability}, 
the BSCC (bottom strongly connected component) decomposition technique was extended to quantum Markov chains and applied in their reachability analysis. {\color{black}The same decomposition was also presented in the field of quantum probability ~\cite{umanita2006classification}  with the name ``the recurrent decomposition''.} Meanwhile, another decomposition, {\color{black}namely}  stationary coherence decomposition of quantum Markov chains, was developed in~\cite{baumgartner2012structures} {\color{black}and further  generalized into the infinite dimensional case \cite{carbone2016irreducible}}. 

This paper continues our studies of quantum Markov chains in \cite{ying2013reachability} and further {\color{black}develops} decomposition techniques that can be used in algorithmic analysis and model checking of quantum Markov chains.  

\smallskip\

\textbf{Contributions of this paper}:  More concretely, our main contributions include: \begin{enumerate}\item  {\color{black}We develop a new decomposition technique, namely periodic decomposition, for quantum Markov chains.} 
\item We give several characterizations of limiting states of quantum Markov chains in terms of aperiodicity, irreducibility, and eigenvalues. 
\item The problem of finding a maximum dimensional noiseless subsystem of a quantum communicating system has been studied employed {\color{black}by} C*-algebra~\cite{chiaverini2004realization,choi2006method} and operator error correction~\cite{kribs2005unified}. We present a new algorithm to solve this problem using decomposition techniques of quantum Markov chains.   
\end{enumerate}

\textbf{Organization of this paper}: We review some basic notions and several useful results of quantum Markov chains from previous literature in Section 2. A two-level decomposition of quantum Markov chains combining the BSCC decomposition  in~\cite{ying2013reachability} and the stationary coherence decomposition in \cite{baumgartner2012structures} is presented in section 3. In particular, an algorithm to compute the two-level decomposition of quantum Markov chains is given. In Section 4, we generalize the notions of irreducibility and periodicity for classical Markov chains to quantum Markov chains. It is shown that they coincide with the corresponding notions defined in the previous literature from different perspectives. We then carefully examine the limiting states of irreducible and aperiodic quantum Markov chains. In particular, a periodic decomposition technique of irreducible quantum Markov chains is presented. In Section 5, we consider general quantum Markov chains that may be reducible. A characterization of their limiting states is given in terms of BSCCs, and their structures are analyzed by combining the stationary coherence,  BSCC and periodic decompositions. As an application of decomposition techniques developed in previous sections, Section 6 shows a method of finding a maximum dimensional noiseless subsystem of a quantum communicating system by reducing the problem to the search of certain BSCCs. 

\section{Quantum Markov Chains}
For convenience of the reader, we review some basic notions and results of quantum Markov chains; for details we refer to \cite{nielsen2010quantum}.
Recall that a (classical) Markov chain is a random process of which the future behavior depends only on the present, and the evolution of such a process is modeled by a matrix of transition probabilities. Note that the evolution of an open quantum system can be modeled mathematically by a \emph{super-operator}, i.e. a completely positive and trace-preserving (CPTP) linear map, acting on its state Hilbert space of the system. This naturally motivates us to {\color{black}present} the following: 
\begin{definition}[\cite{ying2013reachability}] 
A \emph{quantum Markov chain} is a pair $(\mathcal{H}, \mathcal{E})$, where $\mathcal{H}$ is a Hilbert space, and $\mathcal{E}$ is a super-operator on $\mathcal{H}$. \end{definition}

In this paper, we only consider finite-dimensional quantum Markov chains, i.e. dim$(\h)<\infty$. We use $B(\H)$ and $D(\mathcal{H})$ to denote the linear operators acting on $\H$ and  the set of density operators in $\mathcal{H}$, respectively. Then  a \emph{state} of a quantum Markov chain $(\mathcal{H}, \mathcal{E})$ is an operator $\rho\in D(\mathcal{H})$. Recall that the \emph{support} of a density operator $\rho$, denoted $\textrm{supp}(\rho)$, is  the subspace of $\mathcal{H}$ spanned by the eigenvectors of $\rho$ corresponding to non-zero eigenvalues. The \emph{image} of a subspace $\x$ of $\mathcal{H}$ under a super-operator $\mathcal{E}$ is defined to be
\begin{eqnarray*}
\mathcal{E}(\x):=\bigvee_{|\phi\rangle\in \x}\textrm{supp}(\mathcal{E}(|\phi\rangle\langle \phi |)).
\end{eqnarray*}
Here $|\phi\rangle$ denotes a pure state in $\x${\color{black}. The} join of a family $\{\x_{k}\}$ of subspaces of $\mathcal{H}$ is defined by $\bigvee_{k}\x_{k}=\textrm{span}(\bigcup_k \x_{k})$, and for a set of vectors $\y$, $\textrm{span}(\y)=\{\sum_{i=1}^{k}\lambda_{i}v_{i}\mid k\in \mathbb{N}, v_{i}\in \y,\lambda_{i}\in \mathbb{C}\}$
is the space spanned by vectors in $\x$. 

For any linear map $\e$ on $B(\H)$, if $\textrm{dim}(\H)=n$, then it admits up to $n^2$ distinct (complex) eigenvalues $\lambda$  satisfying
\begin{eqnarray*}
  \e(A)=\lambda A
\end{eqnarray*}
for some $A\in B(\H)$, $A\not =0$. We write $\textrm{spec}(\e)$ for the set of all eigenvalues of $\e$. The \emph{spectral radius} of $\e$ is defined as $\varrho(\e):=\sup\{|\lambda| : \lambda \in \textrm{spec}(\e)\}$. In particular, if $\e$ is a CPTP map, then $\varrho(\e)=1$.
\begin{definition}[\cite{yu2012reachability}]\label{reachable-space}
Let $\mathcal{G}=(\mathcal{H}, \mathcal{E})$ be a quantum Markov chain. For any $\rho\in D(\mathcal{H})$, the \emph{reachable space} of $\rho$ is defined to be
$$\mathcal{R}_{\mathcal{G}}(\rho):=\bigvee _{i=0}^{\infty}\textrm{supp}(\mathcal{E}^{i}(\rho))$$ where $\mathcal{E}^{i}$ stands for the composition of $i$ copies of $\mathcal{E}$, that is, $\mathcal{E}^{0}=\mathcal{I}$, the identity super-operator on $\mathcal{H}$, and $\mathcal{E}^{i}=\mathcal{E}^{i-1}\circ\mathcal{E}$ for $i\geq 1$.
\end{definition}

Intuitively, as its name suggests, $\mathcal{R}_{\mathcal{G}}(\rho)$ consists of all states that can be reached from the initial state $\rho$ in the iterative evolution of the system modeled by $\mathcal{G}$. 

\begin{lemma}[\cite{yu2012reachability}]\label{basic2}
Let $\mathcal{G}=(\mathcal{H}, \mathcal{E})$ be a quantum Markov chain and $n=\textrm{dim}(\mathcal{H})$. Then for any state $\rho\in D(\mathcal{H})$, we have
\begin{eqnarray*}
\mathcal{R}_{\mathcal{G}}(\rho)=\bigvee_{i=0}^{n-1}\textrm{supp}(\mathcal{E}^{i}(\rho)).
\end{eqnarray*}
\end{lemma}

The above lemma indicates that all reachable states can be actually reached within $n$ steps if $\textrm{dim}(\mathcal{H})=n$. 
\begin{definition}[\cite{ying2013reachability}]
For a  quantum Markov chain $\mathcal{G}=(\mathcal{H}, \mathcal{E})$, a state $\rho\in D(\mathcal{H})$ is called \emph{stationary} if $\mathcal{E}(\rho)=\rho$; that is, $\rho$ is a fixed point of $\e$. Furthermore, $\rho$ is said to be \emph{minimal} if there is no stationary state $\sigma$ such that $\supp(\sigma)\subsetneq \supp(\rho).$
\end{definition}

Let $\e$ be a super-operator with Kraus operators $\{E_i\}$, i.e. $\e(\cdot)=\sum_i E_i\cdot E_i^\dagger$. Then its matrix representation is defined to be $M=\sum_{i}E_{i}\otimes E_{i}^{*}$ where $E^*$ stands for the (entry-wise) complex conjugate of $E$. Assume that $M=SJS^{-1}$ is the Jordan decomposition of $M$, where
\begin{eqnarray*}
J=\sum_{k=1}^{K}\lambda_{k}P_k+N_k,
\end{eqnarray*}
$N_{k}^{d_k}=0$ for some $d_k>0$, $N_{k}P_{k}=P_{k}N_{k}=N_{k}, P_{k}P_{l}=\delta_{kl}P_{k}, \textrm{tr}(P_{k})=d_k$, and  $\sum_{k}P_{k}=I$. Let
\begin{eqnarray}
\label{equ-1}J_{\infty}& := &\sum_{k:\lambda_{k}=1}P_{k}, \label{eq:infty}\\
\label{equ-2}J_{\phi}& := &\sum_{k:|\lambda_{k}|=1}P_{k}. \label{eq:ephi} 
\end{eqnarray}
Then we write: 
\begin{itemize}
\item $\e_\infty$ for the super-operator with the matrix representation $SJ_{\infty} S^{-1}$.
\item $\e_\phi $ for the super-operator with the matrix representation $SJ_{\phi} S^{-1}$.
\end{itemize}

The following interesting characterizations of $\e_{\infty}$ and $\e_\phi$ are crucial for our later discussion. 

\begin{lemma}[{\cite[Proposition 6.3]{wolf2012quantum}}]\label{lem:inftyphi}
For any quantum Markov chain $\mathcal{G}=(\mathcal{H}, \mathcal{E})$,
\begin{enumerate}
\item  there exists an increasing sequence of integers $n_i$ such that $\e_\phi = \lim_{i\rightarrow \infty} \e^{n_i}$;
\item  $\e_\infty = \lim_{N\rightarrow \infty} \frac 1N \sum_{n=1}^N \e^{n}$.
\end{enumerate}
\end{lemma}

Meanwhile, we collect some other results in the previous literature for later use.
\begin{lemma}[\cite{yu2012reachability}]\label{basic1}
For any quantum Markov chain $\mathcal{G}=(\mathcal{H}, \mathcal{E})$, real number $p>0$, $\rho\in D(\mathcal{H})$, and $\x, \y$ being subspaces of $\mathcal{H}$, we have
\begin{enumerate}
\item  $\textrm{supp}(p\rho)=\textrm{supp}(\rho)$;

\item $\mathcal{E}(\textrm{supp}(\rho))=\textrm{supp}(\mathcal{E}(\rho))$;

\item if $\x\subseteq \y$, then $\mathcal{E}(\x)\subseteq \mathcal{E}(\y)$;
\item  $\mathcal{E}(\x\bigvee \y)=\mathcal{E}(\x)\bigvee \mathcal{E}(\y)$.
\end{enumerate}
\end{lemma}
\begin{lemma}[\cite{paulsen2002completely}]\label{R}
If $\f$ is a CP map with $\f(\cdot)=\sum_{i}F_{i}\cdot F_{i}^{\dagger}$ and $\sum_{i}F_{i}^{\dagger}F_{i}\leq I$, then for any Hermitian matrix $A$,
$$
\| \f(A)\|_{1}\leq \|A\|_{1}{\color{black}.}
$$
\end{lemma}

\begin{lemma}[\cite{horn2012matrix}]\label{minspace}
Let $S(\mathcal{H})$ denote the set of all subspaces of $\mathcal{H}$. Then for any $\rho\in D(\mathcal{H})$, $$\inf_{\x\in S(\mathcal{H})\backslash\{\mathbf{0}\}}\textrm{tr}(P_{\x}\rho)=\lambda_{min}(\rho)$$ where $\mathbf{0}$ is the zero-dimensional subspace and $\lambda_{min}(\rho)$ is the minimum eigenvalue of $\rho$.
\end{lemma}
\section{Two-level Decomposition}
In this section, we show how to combine the BSCC and stationary coherence decompositions developed in {\color{black}\cite{umanita2006classification,ying2013reachability,baumgartner2012structures}} to form a two-level decomposition of quantum Markov chains. In particular, we present an algorithm to implement such a two-level decomposition. 

A central concept in the analysis of quantum Markov chains is the \emph{strongly connected component}. Before giving its definition, let us first introduce an auxiliary notation. Let $\x$ be a subspace of a Hilbert space $\mathcal{H}$, and $\mathcal{E}$ be a super-operator on $\mathcal{H}$. Then the restriction of $\mathcal{E}$ on $\x$ is defined to be a super-operator $\mathcal{E}|_{\x}$ with $\mathcal{E}|_{\x}(\rho)=P_{\x}\mathcal{E}(\rho)P_{\x}$ for all $\rho \in D(\x)$, where $P_{\x}$ is the projector onto $\x$.

\begin{definition}[\cite{ying2013reachability}] 
Let $\mathcal{G}=(\mathcal{H},\mathcal{E})$ be a quantum Markov chain. A subspace $\x$ of $\mathcal{H}$ is called \emph{strongly connected} in $\mathcal{G}$ if for any $|\phi\rangle,|\psi\rangle\in \x$, we have $|\phi\rangle\in \mathcal{R}_{\mathcal{G}_{\x}}(|\psi\rangle\langle \psi |)$ and $|\psi\rangle\in \mathcal{R}_{\mathcal{G}_{\x}}(|\phi\rangle\langle \phi |)$, where $\mathcal{G}_{\x}$ denotes the quantum Markov chain $(\x,\e|_{\x})$; that is, $|\phi\rangle$ and $|\psi\rangle$ can be reached from each other. 
\end{definition}

Let $\SC(\mathcal{G})$ be the set of all strongly connected subspaces of $\mathcal{H}$ in $\mathcal{G}$. It is easy to see that the partial order $(\SC(\mathcal{G}), \subseteq)$ is inductive. Then Zorn's lemma asserts that it has maximal elements. Each maximal element of $(\SC(\mathcal{G}), \subseteq)$ is called a \emph{strongly connected component (SCC)} of $\mathcal{G}$.

\begin{definition}[\cite{ying2013reachability}]\label{BSCC-def}
Let $\mathcal{G}=(\mathcal{H},\mathcal{E})$ be a quantum Markov chain. Then a subspace $\x$ of $\mathcal{H}$ is called a \emph{bottom strongly connected component (BSCC)} of $\mathcal{G}$ if it is a SCC and invariant in $\mathcal{G}$. Here $\x$ is said to be invariant in $\g$ if $\mathcal{E}(\x)\subseteq \x$. 
\end{definition}

The following characterization of BSCCs establishes a connection between BSCCs and minimal stationary states.

\begin{lemma}[\cite{ying2013reachability}]\label{BSCCstationarystate}
	A subspace $\x$ is a BSCC of quantum Markov chain $\g=(\h,\e)$ if and only if there exists a minimal stationary state $\rho^*$ of $\e$ such that supp$(\rho^*)=\x$.
\end{lemma}

Note that a BSCC is essentially equivalent to be a \emph{minimal stationary range} defined in~\cite{baumgartner2012structures}. The following key definition is also from~\cite{baumgartner2012structures}. 
\begin{definition}[\cite{baumgartner2012structures}]\label{SCD}
  Two mutually orthogonal  BSCCs $\b_1$ and $\b_2$ in a quantum Markov chain  $\g=(\h,\e)$ have a \emph{stationary coherence} if there is a BSCC $\b_3\subseteq \b_{1}\bigoplus \b_2$ with $\b_{3}\not =\b_1, \b_{2}$, where $\bigoplus$ denotes direct sum. 
\end{definition}

Stationary coherence is the unique nature of quantum Markov chains, without a counter-part in classical Markov chains. 
\begin{definition}[\cite{ying2013reachability}]\label{transient}
Let $\mathcal{G}=(\mathcal{H},\mathcal{E})$ be a quantum Markov chain. A subspace $\x$ of $\mathcal{H}$ is called a \emph{transient subspace} if for any $\rho\in D(\mathcal{H})$,
\begin{equation}\label{def-transient}
\lim_{n\rightarrow \infty}\textrm{tr}(P_{\x}\mathcal{E}^{n}(\rho))=0{\color{black}.}
\end{equation}
\end{definition}

Intuitively, equation (\ref{def-transient}) means that the system will eventually go out of $\x$ no matter where it starts from. 

Now we are ready to combine the BSCC decomposition in \cite{ying2013reachability} and the stationary coherence decomposition in \cite{baumgartner2012structures} to give a two-level decomposition of quantum Markov chains.  

\begin{theorem}[Two-Level Decomposition]\label{BSCCD}
For any quantum Markov chain $\mathcal{G}=(\mathcal{H}, \mathcal{E})$, we have a unique orthogonal decomposition:
\begin{equation}\label{eqscdecom}
   \h=\bigoplus_{l} \mathcal{X}_{l}\bigoplus \t_{\mathcal{E}}
 \end{equation} 
 where:\begin{enumerate}\item $\t_{\mathcal{E}}$ is the largest transient subspace of $\mathcal{G}.$
\item Each $\mathcal{X}_l$ is either a BSCC or can be further decomposed into mutually orthogonal BSCCs:
\begin{equation}\label{bsccdecom}
\mathcal{X}_l=\b_{l,0}\bigoplus\cdots \bigoplus \b_{l,n_l-1}
\end{equation} such that \begin{itemize} 
\item all BSCCs $\b_{l,j}$ ($0\leq j< n_l$) have the same dimension; and \item there are stationary coherences between any two of them.
\end{itemize}
 \item There are no stationary coherences between BSCCs in $\mathcal{X}_i$ and  $\mathcal{X}_j$ if $i\not = j.$ 
\end{enumerate}
\end{theorem}

From \cite{ying2013reachability}, we know that unlike classical Markov chains, the BSCC decomposition is not unique in general for quantum Markov chains. By Theorem \ref{BSCCD}, we see that the stationary coherence between BSCCs is responsible for this non-uniqueness.
\begin{algorithm}
\caption{Decompose($\g$)}
\label{Irreducibility}
\begin{algorithmic}
\REQUIRE A quantum Markov chain $\mathcal{G}=(\mathcal{H}, \mathcal{E}).$\\
\ENSURE The two-level decomposition of $\h$ in the form of Eqs.(\ref{eqscdecom}) and (\ref{bsccdecom}). \\
$\bigoplus_{i=1}^{n}\b_{i}\bigoplus \t_{\e}$ $\leftarrow$ the BSCC decomposition of $\mathcal{G}=(\mathcal{H}, \mathcal{E})$\\
$I\leftarrow \{1,2,\cdots,n\}$\\
$m\leftarrow 1$\\
\FOR{each $i\in \{1,2,\cdots,n\}$ }
\IF{$i\in I$}
\STATE $x_{m}\leftarrow \{{i}\}$
\FOR{each $j>i$ and $j\in I$}
\STATE$\f\leftarrow\{X\in B(\b_{i}\bigoplus \b_{j}) | \e(X)=X\}$
\IF{dim($\f$) $>2$}
\STATE $x_{m}\leftarrow x_{m}\cup \{{j}\}$
\STATE $I \leftarrow I\backslash\{j\}$
\ENDIF
\ENDFOR
\STATE $\x_m\leftarrow \bigoplus_{k\in x_{m}}\b_{k}$
\STATE $m\leftarrow m+1$
\ENDIF
\ENDFOR
\RETURN $\t_{\e}, \x_i,\{\b_{k}\}_{k\in x_i}$ for $1\leq i\leq m-1$ 
\end{algorithmic}
\end{algorithm}

For practical applications, an algorithm to implement the two-level decomposition is crucial. Actually, we have already an algorithm to compute the BSCC decomposition in \cite{ying2013reachability}. So, a key to computing the two-level composition is to identify the BSCCs having stationary coherences. This problem is solved by the following:

\begin{lemma}\label{decoherence_identify}
  Let $\b_1$ and $\b_2$ be two mutually orthogonal BSCCs of a quantum Markov chain $(\h,\e)$ and $\f=\{X\in B(\b_{1}\bigoplus \b_{2}) | \e(X)=X\}$. Then there is a stationary coherence between $\b_1$ and $\b_2$ if and only if dim$(\f)>2$.
 \end{lemma} 

 {\it Proof.}
By Lemma~\ref{BSCCstationarystate}, $\b\subseteq \h$ is a BSCC  if and only if there is a minimal stationary state $\rho$ such that supp$(\rho)=\b$. If $\b_1$ and $\b_2$ are two mutually orthogonal BSCCs, let $\rho_{1}\in D(\b_1)$ and $\rho_{2}\in D(\b_2)$ be corresponding minimal stationary states. Then there is a  stationary coherence if and only if there is at  least one stationary state $\rho_{3}$  which cannot be linearly represented by $\rho_{1}$ and $\rho_2$. The lemma then follows from \cite[Corollary 6.5]{wolf2012quantum}.
\hfill $\Box$

We further observe that the stationary coherence is transitive in the sense that if there are stationary coherences between $\b_1$ and $\b_2$ and between $\b_2$ and $\b_3$, then there is a stationary coherence between $\b_1$ and $\b_3$ as well.  

No algorithm for computing the stationary coherence decomposition has been given in the previous literature. We developed an algorithm for this purpose. It can be combined with the BSCC decomposition algorithm in \cite{ying2013reachability} to compute 
the two-level decomposition of quantum Markov chains. This combination is presented as Algorithm 1. 
The time complexity of Algorithm 1 is $O(n^8)$, where dim$(\h)=n.$

\section{Irreducibility and Periodicity}
A two-level decomposition of quantum Markov chains was developed in the last section. In this section, we propose another decomposition technique, namely periodic decomposition, which can be combined with the two-level decomposition to further form a three-level decomposition. Such a three-level decomposition provides us with  a very useful tool for a finer algorithmic analysis of quantum Markov chains. 

We first extend the notion of irreducibility for quantum Markov chains{\color{black}, which turns out to be equivalent to the irreducibility defined in the previous literature}. Recall from classical probability theory that an irreducible Markov chain starting from a state can reach any other state in a finite number of steps. With the help of {\color{black}the} reachable space introduced in Definition \ref{reachable-space}, we have: 
\begin{definition}\label{Irreducibility_Def}
A quantum Markov chain $\mathcal{G}=(\mathcal{H}, \mathcal{E})$ is called \emph{irreducible} if for any $\rho\in D(\mathcal{H})$, $\mathcal{R}_{\mathcal{G}}(\rho)=\mathcal{H}$.
\end{definition}

From Lemma~\ref{basic2}, it can be easily shown that the above definition indeed coincides with the irreducibility given in~{\color{black}\cite{davies1970quantum} for quantum stochastic processes and }\cite[Theorem 6.2]{wolf2012quantum} for quantum channels.
However, our definition presents a more natural extension of irreducibility for classical Markov chains.

To illustrate irreducibility, let us see two simple examples.
\begin{example}
\label{reducible}
Consider a natural way to encode the classical NOT gate $X: 0\rightarrow 1; 1\rightarrow 0$ into a quantum operation. Let $\mathcal{H}=\textrm{span}\{|0\rangle, |1\rangle\}$. The
super-operator $\mathcal{E}: D(\mathcal{H}) \rightarrow D(\mathcal{H})$ is defined by 
\begin{eqnarray*}
\mathcal{E}(\rho)=|1\rangle\langle 0|\rho|0\rangle \langle 1|+|0\rangle\langle 1|\rho|1\rangle \langle 0|
\end{eqnarray*}
for any $\rho \in D(\mathcal{H})$. It is easy to check that the quantum Markov chain $(\mathcal{H},\mathcal{E})$ is irreducible.
\end{example}

\begin{example}[Amplitude-damping channel]
Consider the 2-dimensional amplitude-damping channel modeling the physical processes such as spontaneous emission. Let  $\mathcal{H}=\textrm{span}\{|0\rangle, |1\rangle\}$, and
\begin{eqnarray*}
\mathcal{E}(\rho)=E_0 \rho E_0^\dag + E_1 \rho E_1^\dag
\end{eqnarray*}
where $E_{0}=|0\rangle\langle 0|+\sqrt{1-p}|1\rangle\langle 1|$ and $E_{1}=\sqrt{p} |0\rangle\langle 1|$ with $p>0$.
Then the quantum Markov chain $\mathcal{G}=(\mathcal{H}, \mathcal{E})$ is reducible since, say,  $\mathcal{R}_{\mathcal{G}}(|0\rangle\langle 0|) = \textrm{span}\{|0\rangle\}$.
\end{example}

Let us now consider how to check whether a quantum Markov chain $\mathcal{G}=(\mathcal{H}, \mathcal{E})$ is irreducible. Note 
from~\cite[Lemma 3]{ying2013reachability} that $\G$ is irreducible if and only if the state Hilbert space $\H$ itself is a BSCC of $\G$. Moreover, we have the following: 

\begin{lemma}\label{Unique_BSCC}
     A quantum Markov chain $\mathcal{G}=(\mathcal{H}, \mathcal{E})$ has a unique BSCC $\b$ if and only if it has a unique stationary state $\rho^*$. Furthermore, $\textrm{supp}(\rho^*)=\b$.  
\end{lemma}

{\it Proof.}
We see from Definition \ref{transient} that for any stationary state $\rho$, $\textrm{supp}(\rho)\subseteq \b$. Then the result immediately follows from Lemma~\ref{BSCCstationarystate} and Theorem~\ref{BSCCD}. \hfill $\Box$

Therefore, uniqueness of BSCCs in $\G$ can be used to check irreducibility of $\G$. 

\begin{theorem}{\color{black}\cite[Theorem 13]{davies1970quantum}}\label{irreducibility}
A quantum Markov chain $\mathcal{G}=(\mathcal{H}, \mathcal{E})$ is irreducible if and only if it has a unique stationary state $\rho^*$ with $\textrm{supp}(\rho^*)=\mathcal{H}$.
\end{theorem} 

Several different versions of this theorem and its special cases are known in \cite{wolf2012quantum} and \cite{fagnola2009irreducible}. But the above version can be more conveniently used in checking irreducibility of quantum Markov chains. Indeed, it shows that checking whether $\G=(\H,\E)$ is reducible can be done by Algorithm~1 in \cite{ying2013reachability} to check whether its state space $\H$ is a BSCC. The time complexity is $O(n^{6})$, where $\textrm{dim}(\H)=n$.

Next, {\color{black}we consider the periodicity of quantum Markov chains.}
\begin{definition}\label{Def_aperiodic}
Let $\mathcal{G}=(\mathcal{H}, \mathcal{E})$ be a quantum Markov chain. \begin{enumerate}\item A state $\rho \in D(\H)$ is called \emph{aperiodic} if $$\textrm{gcd}\{m\geq 1:\textrm{supp}(\rho) \subseteq \textrm{supp}(\mathcal{E}^{m}(\rho))\}=1.$$
Here, $\textrm{gcd}$ stands for the greatest common divisor; in particular, we assume that $\textrm{gcd}(\emptyset)=0$.

\item A subspace $\x$ of $\H$ is aperiodic if each density operator $\rho$ with $\textrm{supp}(\rho)\subseteq \x$ is aperiodic.

\item If there exists an integer $d\geq1$ such that the whole state space $\H$ is aperiodic in quantum Markov chain $\G^d=(\H,\e^d)$, then the minimum of such integers $d$, denoted $d(\g)$, is called the \emph{period} of $\G$. \item When $d(\g)=1$, $\G$ is said to be \emph{aperiodic}; otherwise, it is \emph{periodic}.\end{enumerate}
\end{definition}

For the special case of irreducible quantum Markov chains, periodicity was defined in {\color{black}\cite{fagnola2009irreducible,carbone2016open}} based on the notion of $\e$-cyclic resolution:
\begin{definition}[\cite{carbone2016open}]\label{ecycle}
For a quantum Markov chain $\G=(\H,\E)$, let $(P_{0},\cdots,P_{d-1})$ be a resolution of identity, i.e. a family of orthogonal projectors such that $\sum_{k=0}^{d-1}P_{k}=I$. Then $(P_{0},\cdots,P_{d-1})$ is said to be $\e$-cyclic if $\e^{\dagger}(P_{k})=P_{k\boxminus1}$ for $k=0,\cdots,d-1$, where  $\boxminus$ denotes subtraction modulo $d$ and $\e^\dagger$ is the adjoint map of $\e$; that is, the linear map such that $\textrm{tr}(M\e(A)) = \textrm{tr}(\e^\dag(M)A)$ for all $M$ and $A$ in $B(\h)$.
\end{definition}

The next lemma shows that the period defined in \cite{carbone2016open} and that in Definition \ref{Def_aperiodic} are the same for irreducible quantum Markov chains. Actually, the former can be better understood in the Heisenberg picture, and the latter in the Schr\"{o}dinger picture. 

\begin{lemma}\label{equivalence}
For an irreducible quantum Markov chain $\G=(\H,\E)$, the period of $\G$ is equal to 
the maximum integer $c$ for which there exists a $\e$-cyclic resolution $(P_{0},\cdots, P_{c-1})$ of identity.
\end{lemma}

{\it Proof.} 
By \cite[Theorem 6.6]{wolf2012quantum}, the maximum $c$ for which there exists a
$\e$-cyclic resolution $(P_{0},\cdots, P_{c-1})$ of identity is the number of the eigenvalues of $\e$ with magnitude one. Then the result follows from Lemma \ref{period}.
\hfill $\Box$

The notion of periodicity is further illustrated by the following example. 
\begin{example}
Let $\mathcal{G}=(\mathcal{H},\mathcal{E})$ with $\mathcal{H}= \textrm {span}\{|0\rangle, |1\rangle, |2\rangle\}$ and
for any $\rho\in D(\H)$,
\begin{eqnarray*}
\mathcal{E}(\rho)=|1+2\rangle\langle 0| \rho |0\rangle\langle 1+2 |+|0+2\rangle\langle 1| \rho |1\rangle\langle 0+2 |
+|1+0\rangle\langle 2| \rho |2\rangle\langle 1+0 |
\end{eqnarray*}
where $|i+j\rangle=(|i\rangle +| j\rangle)/\sqrt{2}$ for  $i, j\in\{0,1, 2\}$. Then it is easy to see that $\mathcal{G}$ is irreducible and aperiodic, and has the unique stationary state
$$
\frac 1 3 (|1+2\rangle\langle 1+2|+|0+2\rangle\langle 0+2|+|1+0\rangle\langle 1+0|).
$$
\end{example}

The following lemma presents a useful characterization of the reachable space starting from a state within an aperiodic subspace. It can be seen as a strengthened version of Lemma~\ref{basic2} in the special case of aperiodic quantum Markov chains. 
\begin{lemma}\label{aperiodic_states}
Let $\mathcal{G}=(\mathcal{H}, \mathcal{E})$ be a quantum Markov chain and $\x$ be an subspace of $\H$. Then the following statements are equivalent:

(1) $\x$ is an aperiodic subspace of $\H$;

(2) For any  $\rho\in D(\mathcal{H})$ with $\textrm{supp}(\rho)\subseteq \x$, there exists an integer $M(\rho)>0$ such that $\textrm{supp}(\mathcal{E}^{m}(\rho))=\R_\G(\rho)$ for all $m\geq M(\rho)$.
\end{lemma}

{\it Proof.} 
$(2)\Rightarrow (1)$ is obvious. So, we only need to show that $(1)\Rightarrow (2)$. Fix an arbitrary $\rho$ with $\textrm{supp}(\rho)\subseteq \x$. For each $i\geq 0$, let $\x_i = \textrm{supp}(\e^i(\rho))$. In particular, $\x_0 = \textrm{supp}(\rho)$. Let $T_\rho = \{i\geq 1: \x_i \supseteq \x_0\}.$
 Then from Lemma~\ref{basic1}, we have: for any $i, j\geq 0$,
\begin{align}
& \x_{i+j} = \e^i(\x_j); \mbox{ and}\label{eq:tmp1}\\
& \mbox{if $i, j\in T_\rho$, then $i+j \in T_\rho$} \label{eq:tmp2}.
\end{align}
By the assumption that $\x$ is aperiodic, we have  $\textrm{gcd}(T_\rho)=1$. Then from \cite{levin2009markov}, there is a finite subset $\{m_{k}\}_{k\in K}$ of $T_\rho$, $\textrm{gcd}\{m_{k}\}_{k\in K}=1$, and an integer $M'(\rho)>0$ such that for any $i \geq M'(\rho)$, there exist positive integers $\{a_{k}\}_{k\in K}$ such that $i=\sum_{k\in K}a_{k}m_{k}$. Thus $i \in T_\rho$ from Eq.~(\ref{eq:tmp2}).

Now let $M(\rho)=M'(\rho)+n-1$ where $n=\textrm{dim}(\H)$, and take any $m\geq M(\rho)$. For all $0\leq i\leq n-1$, we have shown that $m-i\in T_\rho$; that is, 
$\x_{m-i}\supseteq \x_0$. Thus $\x_m\supseteq \x_i$ from Eq.~(\ref{eq:tmp1}), and $\x_m \supseteq 
\mathcal{R}_{\mathcal{G}}(\rho)$ from Lemma~\ref{basic2}. 
Therefore, $\x_m = 
\mathcal{R}_{\mathcal{G}}(\rho)$, as the reverse inclusion trivially holds.\hfill $\Box$

Combining the above lemma with Definition \ref{Irreducibility_Def}, we have: 

\begin{corollary}\label{somen}
Let $\mathcal{G}=(\mathcal{H}, \mathcal{E})$ be an irreducible and aperiodic quantum Markov chain. Then for any $\rho\in D(\mathcal{H})$, there exists an integer $M(\rho)>0$ such that $\textrm{supp}(\mathcal{E}^{m}(\rho))=\mathcal{H}$ for all $m\geq M(\rho)$.
\end{corollary}

The above corollary shows that starting from any state $\rho$, an irreducible and aperiodic quantum Markov chain can reach the whole state space after a finite number of steps. Then it is interesting to see when the whole space can be reached for the first time.
\begin{definition}
Let $\mathcal{G}=(\mathcal{H}, \mathcal{E})$ be an irreducible and aperiodic quantum Markov chain. For each $\rho\in D(\mathcal{H})$, the \emph{saturation time}  of $\rho$ is defined to be 
$$s(\rho)=\inf\{n\geq 1\mid \textrm{supp}(\mathcal{E}^{n}(\rho))=\mathcal{H}\}.$$\end{definition}

It is clear from Corollary \ref{somen} that the infimum in the defining equation of $s(\rho)$ can always be attained. Furthermore, we can show that for an irreducible and aperiodic quantum Markov chain, the saturation time for any initial state has a \emph{universal} upper bound.

\begin{lemma}\label{subonen}
  Let $\mathcal{G}=(\mathcal{H}, \mathcal{E})$ be a quantum Markov chain and $\x$ be an invariant subspace of $\H$. Then the following statements are equivalent:
\begin{enumerate}\item $\mathcal{G}_\x=(\x,\e|_{\x})$  is irreducible and aperiodic;
\item  There exists an integer $M>0$ such that for all $\rho\in D(\x)$, $\textrm{supp}(\mathcal{E}^{m}(\rho))=\x$ for all $m\geq M$.\end{enumerate}
\end{lemma}

{\it Proof.} 
$(2)\Rightarrow (1)$ is obvious. So, we only need to show that 
$(1)\Rightarrow (2)$. Let $s_{\x}(\rho)$ be the saturation time of $\rho$ in $\G_{\x}$. Then for any $\rho\in D(\x)$, let 
$$B(\rho)=\{\sigma\in D(\x)\mid \|\rho-\sigma\|_{1}<\bar{\lambda}_{min}(\mathcal{E}^{s_\x(\rho)}(\rho))\},$$
where $\|\cdot\|_{1}$ is the trace norm and $\bar{\lambda}_{min}(\rho)$ is the minimum non-zero eigenvalue of $\rho$. Obviously, $B(\rho)$ is an  open set. Then $\{ B(\rho)\}_{\rho\in D(\x)}$ is an open cover of $D(\x)$.
As $D(\x)$ is compact, we can find a finite number of density operators $\{\rho_{i}\}_{i\in J}$ such that
\begin{eqnarray*}
D(X)=\bigcup_{i\in J}B(\rho_{i}).
\end{eqnarray*}
In the following, we show for any $\rho\in D(\x)$ and $\sigma \in B(\rho)$, $\textrm{supp}(\mathcal{E}^{m}(\sigma))=\x$ for all $m\geq s(\rho)$.
Then the theorem holds by taking $M=\max_{i\in J}s_\x(\rho_{i})$.
Let  $\y = \textrm{supp}(\mathcal{E}^{s_\x(\rho)}(\sigma))$, and $P_\y$ be the projector onto $\y$. As $\x$ is invariant, $\y\subseteq \x$. Let $P_{\bar{\y}}=I_\x-P_\y$, where $I_\x$ is the identity operator on $\x$. Then
\begin{eqnarray*}
\textrm{tr}(P_{\bar{\y}}\mathcal{E}^{s_\x(\rho)}(\rho))&=&\|P_{\bar{\y}}\mathcal{E}^{s_\x(\rho)}(\rho)P_{\bar{\y}}\|_1\\&=& 
\|P_{\bar{\y}}(\mathcal{E}^{s_\x(\rho)}(\rho)-\mathcal{E}^{s_\x(\rho)}(\sigma))P_{\bar{\y}}\|_1\\&\leq&
\|\mathcal{E}^{s_\x(\rho)}(\rho)-\mathcal{E}^{s_\x(\rho)}(\sigma)\|_{1}\\
&\leq& \|\rho-\sigma\|_{1}\\
&<&\bar{\lambda}_{min}(\mathcal{E}^{s_\x(\rho)}(\rho)).
\end{eqnarray*}
The first two inequalities follow from Lemma \ref{R}.
By Lemma \ref{minspace}, this is only possible when $\y=\x$, since $\x$ is invariant. In other words, $\textrm{supp}(\mathcal{E}^{s_\x(\rho)}(\sigma))=\x$.
Thus
$\textrm{supp}(\mathcal{E}^{s_\x(\rho)-1}(\sigma))\subseteq \textrm{supp}(\mathcal{E}^{s_\x(\rho)}(\sigma))$, and
$\textrm{supp}(\mathcal{E}^{s_\x(\rho)}(\sigma))\subseteq \textrm{supp}(\mathcal{E}^{s_\x(\rho)+1}(\sigma))$ 
from Lemma \ref{basic1}. So $$\textrm{supp}(\mathcal{E}^{s_\x(\rho)+1}(\sigma)) = \x.$$
By induction, we can show that $\textrm{supp}(\mathcal{E}^{m}(\sigma)) = \x$ for all $m\geq s_\x(\rho)$.
\hfill $\Box$

It is worth noting that the integer $M$ in the above theorem does not depend on state $\rho$. This makes it much stronger than Lemma \ref{aperiodic_states}. Considering the whole state space, we have:
\begin{corollary}\label{onen}
Let $\mathcal{G}=(\mathcal{H}, \mathcal{E})$ be a quantum Markov chain. Then the following statements are equivalent:
\begin{enumerate}\item $\mathcal{G}$  is irreducible and aperiodic;
\item  There exists an integer $M>0$ such that for all $\rho\in D(\mathcal{H})$, $\textrm{supp}(\mathcal{E}^{m}(\rho))=\mathcal{H}$ for all $m\geq M$.\end{enumerate}
\end{corollary}

Note that a (classical) Markov chains described by a stochastic $k$-by-$k$ matrix $P$ is irreducible and aperiodic if and only if there exists an integer $m$ such that  $(P^{m})_{i,j}>0$ for all $i$ and $j$. Then by (classical) Perron-Frobenius theory, we have Wielandt's inequality~\cite{wielandt1950unzerlegbare}: the minimum $m\leq k^2-2k+2$. A quantum Wielandt's inequality was recently proved in~\cite{sanz2010quantum}. As its direct application, we see that the minimal $M$ in Corollary~\ref{onen}  satisfies $M\leq n^4$ where $n = \textrm{dim}(\H)$. Then a limit theorem of quantum Markov chain can be directly obtained by combining Corollary~\ref{onen} and  \cite[Theorem 6.7]{wolf2012quantum}.

\begin{theorem}[Limit Theorem]\label{LimitingState}
Let $\mathcal{G}=(\mathcal{H}, \mathcal{E})$ be a quantum Markov chain. Then the following statements are equivalent:
\begin{enumerate}
  \item $\G$ has  a \emph{limiting state} $\rho^{*}$ with $\textrm{supp}(\rho^{*})=\mathcal{H}$ in the sense that
  \begin{eqnarray*}
\lim_{n\rightarrow \infty}\mathcal{E}^{n}(\rho)=\rho^{*} ,\  \forall \rho\in\mathcal{D}(\mathcal{H}).
\end{eqnarray*}
  \item  $\G$ is irreducible and aperiodic;
  \item  1 is the only eigenvalue of $\e$ with magnitude one and the corresponding eigenvector $\rho^*$ is positive definite.
\end{enumerate}
\end{theorem}

{\it Proof.} Direct from \cite[Theorem 6.7]{wolf2012quantum}, by noting that irreducibility plus aperiodicity are equivalent to primitivity with Corollary~\ref{onen}. \hfill $\Box$

Generally, aperiodicity can be determined by the eigenvalues of $\e$ without the assumption of irreducibility.
\begin{lemma}\label{aperiodic_eigenvalue1}
	Let $\mathcal{G}=(\mathcal{H}, \mathcal{E})$ be a quantum Markov chain with a trivial transient subspace; that is, $\t_\e = \{0\}$ in the decomposition in Eq.~(\ref{bsccdecom}). If  $\e$ has only 1 as its eigenvalue with magnitude one, then $\G$ must be aperiodic. 
\end{lemma}

{\it Proof.} As 1 is the only eigenvalue with magnitude one, $\e_{\phi}=\lim_{n\rightarrow\infty }\e^n$. Then for any $|\psi\rangle \in \h$, 
$$\lim_{n\rightarrow \infty}\e^n(|\psi\rangle\langle \psi |)=\rho^*$$
for some stationary state $\rho^*$.  

By the proof of Lemma \ref{subonen}, there exists an integer $N>0$ such that for all $n>N$, $$\textrm{supp}(\e^n(|\psi\rangle\langle \psi |))=\textrm{supp}(\rho^*).$$ Then with $\t_\e = \{0\}$ and Lemma~\ref{BSCCstationarystate}, there is a stationary state $\sigma^*$ such that $\textrm{tr}(\sigma^* \rho^*)=0$ and $\textrm{supp}(\rho^*+\sigma^*)=\h$. Then as $\textrm{supp}(\rho^*)$, $\textrm{supp}(\sigma^*)$ is invariant under $\e_{\phi}$ and $\e_{\phi}=\lim_{n\rightarrow\infty }\e^n$ is CPTP, it is easy to see that 
$|\psi\rangle \in\textrm{supp}(\rho^*) $. Therefore, by Definition \ref{Def_aperiodic}, $|\psi\rangle\langle \psi |$ is aperiodic. Consequently, $(\H, \e)$ is aperiodic from the arbitrariness of $|\psi\rangle$.\hfill $\Box$

The next lemma shows that the period of an irreducible quantum Markov chain is exactly the number of eigenvalues with magnitude one.
\begin{lemma}\label{period}
  For an irreducible quantum Markov chain $\mathcal{G}=(\mathcal{H}, \mathcal{E})$, the period of $\mathcal{G}$ equals the number of eigenvalues of $\e$ with magnitude one. 
\end{lemma}

{\it Proof.} Let $m$ be the number of eigenvalues of $\e$ with magnitude one and $d$ the period of $\g$. By \cite[Theorem 6.6]{wolf2012quantum}, 1 is the only element in $\textrm{spec}(\e^m)$ with magnitude one. As $\G$ is irreducible,  $\G^m=(\H,\e^m)$ has only a trivial transient subspace by Theorem~\ref{irreducibility}. Then from Lemma \ref{aperiodic_eigenvalue1}, $\G^m=(\H,\e^m)$ is aperiodic, and hence $d\leq m$.

We now turn to prove that $d\geq m$. The case when  $m=1$ is trivial. Suppose $m\geq 2$. By \cite[Theorem 6.6]{wolf2012quantum}, there exists a $\e$-cyclic resolution  of identity $(P_{0},\cdots, P_{m-1})$. As $\G^{d}=(\H,\e^{d})$ is aperiodic, there exists an integer $N'>0$ such that $\textrm{supp}(P_{k})\subseteq$ $\textrm{supp}(\e^{dn}(P_{k}))$ for all $n\geq N'$. Thus for $n\geq N'$
\begin{eqnarray}
  0<\textrm{tr}(P_k\e^{dn}(P_k))=\textrm{tr}(\e^{\dagger^{dn}}(P_k)P_k)=\textrm{tr}(P_{k\boxminus dn}P_k)
\end{eqnarray}
where  $\boxminus$ denotes subtraction  modulo $m$. Therefore, $m$ must be a factor of $d$ and $m\leq d$. \hfill $\Box$

Lemma \ref{period} indicates that every irreducible quantum Markov chain has a period and also offers an efficient algorithm for computing the period by counting the number of eigenvalues of the super-operator. 

Now we are ready to present the main result of this section -- {\color{black}a} periodic decomposition technique for irreducible quantum Markov chains. 

\begin{theorem}[Periodic Decomposition]\label{decomposition_period}
The state Hilbert space $\mathcal{H}$ of an irreducible quantum Markov chain $\mathcal{G}=(\mathcal{H}, \mathcal{E})$ with period $d$ can be decomposed into the direct sum of some orthogonal subspaces:
\begin{eqnarray*}
\mathcal{H}=\b_{0}\bigoplus\cdots \bigoplus \b_{d-1}
\end{eqnarray*}
with the following properties: 
\begin{enumerate}
\item $\e(\b_{i\boxminus1})=\b_{i}$, where $\boxminus$ denotes subtraction modulo $d$;
\item $(\b_{i}, \e^{d}|_{\b_{i}})$ is irreducible and aperiodic; and 
\item $\b_{i}'$s are mutually orthogonal subspaces of $\H$ and invariant under $\e^d$.\end{enumerate}
\end{theorem}

{\it Proof.} Immediate from the proof of  Lemma \ref{period}.\hfill $\Box$

\section{Three-level decomposition for quantum Markov chains}\label{reducible_qMC}

The periodic decomposition technique for irreducible quantum Markov chains was developed in the last section. The main aim of this section is to integrate it with the two-level decomposition to form a finer decomposition of a general quantum Markov chain that might be reducible. 

Let us first consider limiting states in a general quantum Markov chain.  
\begin{lemma}\label{generallimitingstate}
Let $\mathcal{G}=(\mathcal{H}, \mathcal{E})$ be a quantum Markov chain. Then the following statements are equivalent:
\begin{enumerate} 
\item For any $\rho\in D(\mathcal{H})$, $\lim_{n\rightarrow \infty}\mathcal{E}^{n}(\rho)$ exists. 
\item 1 is the only  eigenvalue of $\mathcal{E}$ with magnitude one.\end{enumerate}
\end{lemma}

{\it Proof.}  
If 1 is the only  eigenvalue with magnitude one, then $\mathcal{E}_{\phi}=\lim_{n\rightarrow \infty}\mathcal{E}^{n}$. 
On the other hand, if for any $\rho\in D(\mathcal{H})$, $\lim_{n\rightarrow \infty}\mathcal{E}^{n}(\rho)$ exists, then
\begin{eqnarray*}
\mathcal{E}_{\phi}(\rho)=\lim_{n\rightarrow \infty}\mathcal{E}^{n}(\rho)=\mathcal{E}(\lim_{n\rightarrow \infty}\mathcal{E}^{n}(\rho))=\mathcal{E}(\mathcal{E}_{\phi}(\rho)).
\end{eqnarray*}
So $\mathcal{E}(\mathcal{E}_{\phi})=\mathcal{E}_{\phi}$. Note that the corresponding Jordan norm forms of $\mathcal{E}\circ \mathcal{E}_{\phi}$ and $\mathcal{E}_{\phi}$ are respectively
\begin{eqnarray*}
JJ_{\phi}&=&\sum_{k:|\lambda_{k}|=1}\lambda_{k}P_{k}\\
J_{\phi}&=&\sum_{k:|\lambda_{k}|=1}P_{k}.
\end{eqnarray*}
Thus,  whenever $|\lambda_{k}|=1$ it actually holds $\lambda_{k}=1$.\hfill $\Box$

A special case of Lemma \ref{generallimitingstate} where $\e$ is unital (that is, $\e(I) = I$) was proved in \cite{liu2011limiting}. The following lemma further deals with the case when the limiting state is unique. 
\begin{lemma}\label{limit-theorem}
For any  quantum Markov chain $\mathcal{G}=(\mathcal{H},\mathcal{E})$, the following statements are equivalent:\begin{enumerate}
\item There is a limiting state $\rho^{*}$, i.e. $\lim_{n\rightarrow \infty}\mathcal{E}^{n}(\rho)=\rho^*$ for all $\rho\in D(\mathcal{H})$. Especially, if $\mathcal{E}$ is unital, then $\rho^{*}= I/n$, where dim$(\h)=n$; 
\item $\mathcal{H}$ contains a unique BSCC $\b$ and $(\b,\mathcal{E}|_{\b})$ is aperiodic;
\item $1$ is the only eigenvalue of $\mathcal{E}$ with magnitude one and its geometric multiplicity is 1.
\end{enumerate}

\end{lemma}

{\it Proof.} (1) $\Rightarrow$ (2) is easy. If $\h$ has two BSCCs, then there are two stationary states. This contradicts the uniqueness of limiting states. As limiting states must be stationary, by Lemma \ref{Unique_BSCC}, $\textrm{supp}(\rho^*)=\b$. From Theorem \ref{LimitingState}, $(\b,\mathcal{E}|_{\b})$ is aperiodic.

(2) $\Rightarrow$ (3) As $\e_{\phi}=\lim_{i\rightarrow \infty}\e^{n_i}$, we see from Definition \ref{transient} that for any $\rho \in D(\H)$, $\textrm{supp}(\e_{\phi}(\rho))\subseteq \b$. Note that $(\b,\e|_{\b})$ is irreducible and aperiodic. Thus for any $\sigma \in D(\b)$, $\lim_{n\rightarrow\infty}\e^n(\sigma)=\rho^*$ with $\textrm{supp}(\rho^*)=\b$, and $\e_{\phi}(\sigma)=\rho^*$. From the fact $\e_{\phi}\circ \e_{\phi}=\e_{\phi}$, we have that for any $\rho\in D(\H)$, $\e_{\phi}(\rho)=\rho^*$ and $\rho^*$ is the only stationary state of $\e_{\phi}$. By the definition of $\e_{\phi}$,  the stationary states of $\mathcal{E}$ are also  stationary states of $\mathcal{E}_{\phi}$, so 1 is the only eigenvalue of $\e$ with magnitude one and  its geometric multiplicity is 1. 

(3) $\Rightarrow$ (1) Suppose 1 is the only  eigenvalue with magnitude one. Then $\mathcal{E}_{\phi}=\lim_{n\rightarrow \infty}\mathcal{E}^{n}$, i.e., for all $\rho\in D(\h)$, $\lim_{n\rightarrow\infty}\e(\rho)$ exists. Furthermore, as limiting states must be stationary states and 1's geometric multiplicity is one,  there is only one stationary state $\rho^{*}$ satisfying $\lim_{n\rightarrow\infty}\e(\rho)=\rho^*$ for any $\rho\in D(\h)$, by \cite[Corollary 6.5]{wolf2012quantum}.
\hfill $\Box$

Now we turn to present a three-level decomposition of quantum Markov chains. For any quantum Markov chain $\mathcal{G}=(\mathcal{H},\mathcal{E})$, we first use Theorem \ref{BSCCD} to decompose $\h$ into $$\h=\x_{0}\bigoplus \cdots \bigoplus \x_{n-1}\bigoplus \t_{\e},$$ where  each $\x_l= \b_{l,0}\bigoplus\cdots\bigoplus \b_{l,n_l-1}$. 
It was proved in \cite{ying2013reachability} that although the BSCC decomposition of $\h$ is not unique, the number $n_l$ of BSCCs for different decompositions of $\x_l$ is the same.
Furthermore, for each $\b_{i,l}$, we can employ Theorem \ref{decomposition_period} to decompose it into $d_{i,l}$ aperiodic subspaces, where $d_{i,l}$ is the period of $\b_{i,l}$. Then the only question that remains to answer is: {\color{black}is} the sum $\sum_{i} d_{i,l}$ of the periods of BSCCs the same for different decompositions of $\x_{l}$?  The following is a key lemma to give a positive answer to this question: 
\begin{lemma}\label{lem:sameperiod}
Let $\mathcal{G}=(\mathcal{H},\mathcal{E})$ be a quantum Markov chain with the stationary coherence decomposition{\color{black}:} 
$$\h=\bigoplus_{l}\x_{l}\bigoplus \t_{\e}{\color{black}.}$$
Then for any $l$ and any BSCCs $X$ and $Y$ contained in $\x_l$, we have $d(X) = d(Y)$, where $d(\cdot)$ denotes the period of $\e$ when restricting to the corresponding subspace.
\end{lemma}

{\it Proof.} Let $\r=\bigoplus_{l}\x_l$ be the subspace of $\h$ spanned by all BSCCs. For any subspace $Z$ of $\h$, let $P_Z$ be the projector onto $Z$ and $\p_Z$ be super-operator with $\p_{Z}(\cdot)=P_Z \cdot P_Z$.  Then $\e|_Z = \p_Z \circ \e \circ \p_Z$. By \cite[Corollary 23]{baumgartner2012structures}, we can find a unitary $U$ such that \begin{enumerate}\item $P_Y = UP_XU^\dag$ (thus $\p_Y = \u\circ \p_X\circ\ \u^\dag$ where $\u(\cdot) = U\cdot U^\dag$); and \item for any linear operator $A$ with $A = \p_\r(A)$,
$\p_\r\circ \e^\dag \circ \u(A) = \u\circ \p_\r \circ \e^\dag(A)$.\end{enumerate}

For any orthogonal projectors $P_0, \cdots, P_{d-1}$ such that $\sum_{i=0}^{d-1} P_i = P_X$ and $\e_X^\dag(P_i)  = \p_X\circ \e^\dag(P_i) = P_{i\boxminus 1}$, where $\boxminus$ denotes subtraction modulo $d$, let $P'_0, \cdots, P'_{d-1}$ be orthogonal projectors with $P'_i = \u(P_i)$. Then for any $i$,
\begin{eqnarray*}
(\e|_Y)^\dag(P'_i) & =& \p_Y\circ \e^\dag\circ\p_Y\circ \u(P_i) = \p_Y\circ  \p_\r\circ \e^\dag\circ \u(P_i)\\
& =& \p_Y\circ \u\circ  \p_\r\circ \e^\dag(P_i) = \u\circ \p_X\circ  \p_\r\circ \e^\dag(P_i)\\
& =& \u\circ \p_X\circ \e^\dag(P_i) = \u(P_{i \boxminus 1}) = P'_{i\boxminus 1}.
\end{eqnarray*}
Thus following from Lemma~\ref{equivalence}, $d(X) \leq d(Y)$. By a symmetric argument, we can show that $d(Y) \leq d(X)$ as well.
\hfill $\Box$

Now we can easily prove the following: 
\begin{theorem}\label{cor:unique}
Let a quantum Markov chain $(\h, \e)$ have two different BSCC decompositions: 
\begin{eqnarray*}\H &=&  \b_{0}\bigoplus\cdots\bigoplus \b_{n-1} \bigoplus \t_\e \\
 &=& \b'_{0}\bigoplus\cdots\bigoplus \b'_{n-1}\bigoplus \t_\e\end{eqnarray*}
and let $d_i$ (resp. $d_i'$) be the period of $\e$ restricting on $\b_i$ (resp. $\b_i'$). Then
$$\sum_{i=0}^{n-1} d_i = \sum_{i=0}^{n-1} d_i'.$$
\end{theorem}
\section{An Application}
To show the utility of the three-level decomposition for quantum Markov chains presented in the last section, we apply it to the problem of finding a maximum dimensional 
 noiseless subsystem of a quantum communicating system. The problem has been tackled in the previous literature~\cite{chiaverini2004realization,choi2006method} using C*-algebra and operator error correction. The advantage of our approach is that we can give an algorithmic solution to it. 
 
 In quantum communication, channels are mathematically modeled as super-operators. Therefore, a simple quantum communicating system can be regarded as a quantum Markov chain.
 
\begin{definition}[\cite{kribs2005unified}]
	Given a quantum Markov chain $\G=(\h, \e)$ with $\h=(\h_{A}\bigotimes\h_{B})\bigoplus\mathcal{K}$.\begin{enumerate}\item If dim$(\h_A)>1$, then $\h_{A}$ is called a \emph{subsystem} of $\h$. \item A subsystem $\h_{A}$ is \emph{noiseless} if 
	$$\forall \rho_{A}\in D(\h_A),\rho_{B}\in D(\h_{B})  \ \exists \sigma_{B}\in D(\h_{B}):\e(\rho_{A}\otimes \rho_{B})=\rho_{A}\otimes\sigma_{B}.$$\end{enumerate}
\end{definition}

Intuitively, noiselessness means that quantum operation $\e$ does not cause any change of the state of subsystem $\h_A$. In other words, the restriction of $\e$ onto $\h_A\bigotimes \h_B $ satisfies 
$$
\e_{AB}=I_{A}\otimes \e_{B}
$$ 
for some CPTP map $\e_{B}$ on $\h_{B}$; That is the noisy subsystem $\h_B$ has no importance.

The next lemma shows that to decide if a subsystem $\h_A$ is noiseless with respect to $\e$, it suffices to find a state $\sigma_B$  such that any product state $\rho_A\otimes \sigma_B$ is in $D(\h)$ and  a stationary state. 

\begin{lemma}\label{fixed_point}
	Given a quantum operation $\e$ on $\h$, a subsystem $\h_A$ is noiseless if and only if 
	$$\exists \sigma_B, \ \forall \rho_A\in D(\h_A) : \rho_A\otimes \sigma_B\in D(\h)	\textrm{ and }\e(\rho_A\otimes \sigma_B)=\rho_A\otimes \sigma_B.$$
\end{lemma}

{\it Proof.}
Let $\h=\h_A\bigotimes \h_B\bigoplus \k$ and $P$ be the projector onto  $\h_{A}\bigotimes \h_{B}$. If $\h_A$ is noiseless, then the restriction of $\e$ to $\h_A\bigotimes \h_B $ satisfies 
$$
\e_{AB}=I_{A}\otimes \e_{B}
$$ 
for some CPTP map $\e_{B}$ on $\h_{B}$. So there exists a stationary state $\sigma_B$ for $\e_{B}$ from Theorem~\ref{BSCCD} and Lemma~\ref{BSCCstationarystate}. Then we have that for any $\rho_A\in D(\h_{A})$,
$$
	\e(\rho_A\otimes \sigma_B)=\rho_A\otimes \sigma_B.
	$$

Conversely, let $\h_B=\textrm{supp}(\sigma_B)$. By the assumption, for any pure state $|\psi\rangle\in \h_{A}$, $\e(|\psi\rangle\langle \psi|\otimes \sigma_B)=|\psi\rangle\langle \psi|\otimes \sigma_B $. Then $\e(P)=P\e(P)P\Rightarrow E_{i}P=PE_i P$, where $\{E_{i}\}$ is the Kraus operators of $\e$. Then let $K_i=PE_i P\ \forall i$. 

First, we claim that $(|\psi\rangle\langle \psi|\otimes P_{B})K_{i}=K_{i}(|\psi\rangle\langle \psi|\otimes P_{B}) \ \forall i$ by the similar argument in the proof of \cite[Lemma 5.2]{blume2010information}. With the arbitrariness of $|\psi\rangle$, for any $i$, $K_i$ must have the following form:
$$K_{i}=I_{A}\otimes F_{i}$$
for some operator $F_{i}$ on $\h_{B}$. Therefore, $\h_{A}$ is noiseless.
\hfill $\Box$

Now we are ready to give a sufficient and necessary condition for noiselessness. 

\begin{theorem}\label{BSCCs}
	Given a quantum Markov chain $\G=(\h,\e)$ with a subsystem $\h_A$.  $\h_A$ is noiseless if and only if  there exists a subspace $\h_B$ such that for any $|\psi\rangle\in \h_A$, $\textrm{supp}(|\psi\rangle\langle \psi|)\bigotimes \h_{B}\subseteq \h$ is a BSCC of $\e$.  
\end{theorem}

{\it Proof.} Let $\h=\h_A\bigotimes\h_B\bigoplus\k$. If $\h_A$ is noiseless, then the restriction of $\e$ to $\h_A\bigotimes \h_B $ satisfies 
$$
\e_{AB}=I_{A}\otimes \e_{B}
$$ 
for some CPTP map $\e_{B}$ on $\h_{B}$. Let $\sigma$ be a minimal stationary state of $\e_{B}$ in $\h_B$. Now, we claim that for any $|\psi\rangle \in \h_A$, $|\psi\rangle \langle \psi |\otimes \sigma $ is a minimal stationary state of $\e$. 

Firstly, $\e(|\psi\rangle \langle \psi |\otimes \sigma )=|\psi\rangle \langle \psi |\otimes \sigma $. Thus $\h_{\psi}\bigotimes \h_{B'}$ is an invariant subspace of $\e$, where $\h_{\psi}=\textrm{supp}(|\psi\rangle\langle \psi |)$ and $\h_{B'}=$ supp$(\sigma)$.  If there exists a stationary state $\rho\in D(\h_{\psi}\bigotimes \h_{B'})$ of $\e$ and $\rho\neq |\psi\rangle \langle \psi |\otimes \sigma $, then $\rho$ must have the following form:
$$\rho=|\psi\rangle \langle \psi |\otimes \sigma'$$
for some $\sigma'\in D(\h_{B'})$ and $\sigma'\neq \sigma$. Hence $\sigma '$ is the stationary state of $\e_{B}$ in $\e_{AB}=I_{A}\otimes \e_{B}$, contradicting that $\sigma$ is a minimal stationary state of $\e_{B}$. Then $\h_{\psi}\bigotimes \h_{B'}$ is a BSCC following from Lemma~\ref{BSCCstationarystate}.

Conversely, let $\{|i\rangle\}_{i=1}^{n}$ be a set of orthogonal basis of $\h_{A}$, where $n=$ dim$(\h_A)$. Then we have a BSCC decomposition for $\h_{A}\bigotimes \h_{B}$ as follows:
$$
\h_{A}\bigotimes \h_{B'}=\bigoplus_i (\h_i\bigotimes \h_{B})
$$
where $\h_i=\textrm{supp}(|i\rangle\langle i|)$ for all $i$.
As for all $|\psi\rangle\in \h_A$, $\h_{\psi}\bigotimes \h_{B}$ is a BSCC,  there are stationary coherences between any two BSCCs in the above decomposition. From \cite[Theorem 7]{baumgartner2012structures}, we obtain that:
$$\v_{A}=U_{A}\h_{A}, \ \ \ \v_{B}=U_{B}\h_{B}$$
for some unitary matrices $U_{A}$, $U_{B}$ and $\h_{A}\bigotimes \h_{B}=\v_{A}\bigotimes \v_{B}$, such that there is a state $\tau$ with supp$(\tau)= \v_{B}$ and for any stationary state  $\rho\in D(\h_{A}\bigotimes \h_{B})$ of $\e$,
$$\rho=U_{A}^\dagger\rho'U_{A}\otimes U_{B}^\dagger\tau U_{B}$$
for some $\rho'\in D(\v_{A})$. Note again that for any $|\psi\rangle\in \h_{A}$, $\h_{\psi}\bigotimes\h_{B} $ is a BSCC. There exists a state $\rho_{\psi}$  with supp$(\rho_{\psi})=\h_{B}$ such that $|\psi\rangle\langle \psi |\otimes \rho_{\psi}\in D(\h_{A}\bigotimes \h_{B})$ is stationary under $\e$. Therefore, 
$$|\psi\rangle\langle \psi |\otimes \rho_{\psi}=U_{A}^\dagger|\phi\rangle\langle \phi |U_{A}\otimes U_{B}^\dagger\tau U_{B}$$ for some $|\phi\rangle\in \h_{A}$, which is possible only if $\rho_{\psi}=U_{B}^\dagger\tau U_{B}$. By the arbitrariness of $\psi$ and the linearity of $\e$, we know
$$\e(\rho_A\otimes U_{B}^\dagger\tau U_{B} )=\rho_A\otimes U_{B}^\dagger\tau U_{B}$$ for any $\rho_A\in D(\h_{A})$. Then the result follows from Lemma \ref{fixed_point}.  \hfill $\Box$

Furthermore, the maximum dimension of noiseless subsystems can be determined as follows.
\begin{corollary}
For a quantum operation $\e$ on $\h$, the maximum dimension of noiseless subsystems is the maximum number of BSCCs which have stationary coherences between any two of them. 
\end{corollary}
{\it Proof.} Immediate from \cite[Theorem 7]{baumgartner2012structures} and Theorem~\ref{BSCCs}.\hfill $\Box$
\begin{algorithm}
\caption{FindMaxNS($\g$)}
\label{FindMaxNS}
\begin{algorithmic}
\REQUIRE A quantum Markov chain $\g=(\h,\e)$.\\
\ENSURE A decomposition of $\h=(\h_{A}\bigotimes \h_{B})\bigoplus \k$ where $\h_A$ is a maximum dimensional noiseless subsystem or false indicating there is no noiseless subsystem. \\
$\h=\bigoplus_{l=0}^{m-1}\x_{l}\bigoplus\t_{\e}, \x_{l}=\bigoplus_{i=0}^{n_{l}-1}\b_{i,l}\leftarrow$ Decompose(\g)\\
$f\leftarrow 1$\\
$j\leftarrow 0$
\FOR{each $l\in \{0,1,\cdots,m-1\}$ }
\IF{$f<n_{l}$}
\STATE $f\leftarrow n_{l}$
\STATE $j\leftarrow l$
\ENDIF
\ENDFOR
\IF{$f>1$}
\STATE $\k\leftarrow\bigoplus_{l=0,l\not=j}^{m-1}\x_{l}\bigoplus\t_\e$
\STATE $\h_A\leftarrow \mathbb{C}^{f}$
\STATE $r\leftarrow$ the dimension of $\b_{0,j}$
\STATE $\h_B\leftarrow \mathbb{C}^{r}$
\RETURN $\h_A,\h_B,\mathcal{K}$
\ELSE
\RETURN \FALSE
\ENDIF
\end{algorithmic}
\end{algorithm}

It is clear from Theorem \ref{BSCCs} and \cite[Theorem 7]{baumgartner2012structures} that finding a maximum dimensional noiseless system is equivalent to searching special BSCCs, which have stationary coherences between any two of them. 
So, Algorithm 1 can serve as a footstone of the construction of the noiseless subsystem. In general, the number of noiseless subsystems is infinite for a given quantum communicating system. But we are able to develop an algorithm -- Algorithm 2 -- 
for finding a maximum dimensional noiseless subsystem, which is usually the most important in practical applications. It is easy to see that Algorithm 2 either produces a maximum one or {\color{black}indicates} there is no noiseless subsystem in time $O(n^8)$, where dim$(\h)=n$.

\section{Conclusion}
In this paper, we {\color{black}obtained some useful characterizations of}  irreducibility and periodicity for quantum Markov chains. Based on them, we developed a periodic decomposition technique for irreducible quantum Markov chains, which is further combined with the BSCC decomposition and stationary coherence decomposition in the previous literature to construct a three-level decomposition of general quantum Markov chains. This three-level decomposition provides us with a finer tool for algorithmic analysis and model-checking of quantum systems. We also established a limit theorem that gives a characterization of limiting states in a quantum Markov chain in terms of periodicity, irreducibility, and eigenvalues of the super-operator. As an {\color{black}application}, we presented an algorithm for constructing the maximal dimensional noiseless {\color{black}subsystem} of a quantum communicating system.

There are several interesting topics for future studies:
\begin{itemize}\item Reachability analysis of quantum Markov chains: The BSCC decomposition was already used in reachability analysis of quantum Markov chains \cite{ying2013reachability}. The eventual, global, ultimately forever and infinitely often reachability of quantum automata were carefully examined in \cite{li:reachability}. Quantum automata {\color{black}is} a special kind of quantum Markov chains, where the dynamics is described by a unitary transformation rather than a general super-operator. It seems that the three-level decomposition presented in this paper is useful for analysis of these more sophisticated reachability of quantum Markov chains. 
\item Extend the decomposition techniques developed in this paper to quantum Markov decision processes, which were introduced in \cite{Aaronson} for quantum machine learning and in \cite{ying:decision} for modeling concurrent quantum programs.  
\end{itemize}
\subparagraph*{Acknowledgements.}
This work was partly supported by the Australian Research Council (Grant No: DP160101652), the Key Research Program of Frontier Sciences, Chinese Academy of Sciences, and the CAS/SAFEA International Partnership Program for Creative Research Team.

\section*{References}

\bibliography{bib}

\end{document}